\documentclass[12pt]{article}
\usepackage{amsmath}
\usepackage{graphicx}
\usepackage{enumerate}
\usepackage{natbib}
\usepackage{url} % not crucial - just used below for the URL 

\RequirePackage{amsthm,amsfonts,amssymb}
\usepackage{algorithmic}
\usepackage[ruled,vlined]{algorithm2e}
\usepackage{caption}
\usepackage{subcaption}
\usepackage{cancel}

\newcommand{\blind}{1}

\begin{document}

\def\spacingset#1{\renewcommand{\baselinestretch}%
{#1}\small\normalsize} \spacingset{1}
\addtolength{\textheight}{.5in}%

%%%%%%%%%%%%%%%%%%%%%%%%%%%%%%%%%%%%%%%%%%%%%%%%%%%%%%%%%%%%%%%%%%%%%%%%%%%%%%

\if1\blind
{
  \title{\bf Learning from limited temporal data: Dynamically sparse historical functional linear models with applications to Earth science}
  \author{Joseph Janssen \thanks{
    This research was funded by the Collaborative Research Team Project of Canadian Statistical Sciences Institute (CANSSI) awarded to Ali A. Ameli, William J. Welch, and Jiguo Cao. Joseph Janssen was supported by Natural Sciences and Engineering Research Council of Canada (NSERC) PhD Scholarship. Asad Haris was supported by postdoctoral funding from UBC's Data Science Institute. Jiguo Cao's and William J.\ Welch's research was partially supported by NSERC Discovery grants (RGPIN-2018-06008 and RGPIN-2019-05019, respectively). Stefan Schrunner gratefully acknowledges financial support from the Norwegian University of Life Sciences (project number 1211130114) for an international stay at the University of British Columbia, Canada.
    }\hspace{.2cm}\\
    \footnotesize Department of Earth, Ocean and Atmospheric Sciences, University of British Columbia\\
    and \\
    Shizhe Meng \\
    \footnotesize Department of Mathematics, University of British Columbia\\
    and \\
    Asad Haris \\
    \footnotesize Department of Earth, Ocean and Atmospheric Sciences, University of British Columbia\\
    and \\
    Stefan Schrunner \\
    \footnotesize Department of Data Science, Norwegian University of Life Sciences\\
    and \\
    Jiguo Cao \\
    \footnotesize Department of Statistics and Actuarial Science, Simon Fraser University\\
    and \\
    William J. Welch \\
    \footnotesize Department of Statistics, University of British Columbia\\
    and \\
    Nadja Kunz \\
    \footnotesize Norman B Keevil Institute of Mining Engineering, University of British Columbia\\
    and \\
    Ali A. Ameli \\
    \footnotesize Department of Earth, Ocean and Atmospheric Sciences, University of British Columbia}
  \maketitle
} \fi

\if0\blind
{
  \bigskip
  \bigskip
  \bigskip
  \begin{center}
    {\LARGE\bf Learning from limited temporal data: Dynamically sparse historical functional linear models with applications to Earth science}
\end{center}
  \medskip
} \fi

\newpage
\begin{abstract}
%The text of your abstract. 200 or fewer words.
Scientists and statisticians often want to learn about the complex relationships that connect two time-varying variables. Recent work on sparse functional historical linear models confirms that they are promising for this purpose, but several notable limitations exist. Most importantly, previous works have imposed sparsity on the historical coefficient function, but have not allowed the sparsity, hence lag, to vary with time. We simplify the framework of sparse functional historical linear models by using a rectangular coefficient structure along with Whittaker smoothing, then reduce the assumptions of the previous frameworks by estimating the dynamic time lag from a hierarchical coefficient structure. We motivate our study by aiming to extract the physical rainfall-runoff processes hidden within hydrological data. We show the promise and accuracy of our method using eight simulation studies, further justified by two real sets of hydrological data.
\end{abstract}

\noindent%
{\it Keywords:}  
%3 to 6 keywords, that do not appear in the title
Functional data analysis, Hydrology, Rainfall-runoff relationships, Streamflow, Nonstationary time lag, Unit hydrograph

\vfill

\newpage
\spacingset{1.9} % DON'T change the spacing!
\section{Introduction}
\label{sec:intro}
%%%%% TODO %%%%%%
%% cite Enhancing Sparsity by Reweighted L1 Minimization, VaRIABLE SELECTION AND ESTIMATION IN HIGH-DIMENSIONAL VARYING-COEFFICIENT MODELS, Least absolute shrinkage is equivalent to quadratic penalization, On the use of cross-validation for the calibration of the adaptive lasso, An Adaptive Ridge Procedure for L0 Regularization, Broken adaptive ridge regression and its asymptotic properties

% Not doing forecast, hot-moment of celerity+ explain (in which day of they year will thunderstorm cause flood/ or cause nothing)

In many complex systems, a causal variable is filtered into some response of interest via complicated and unknown non-stationary processes. Exactly how the causal impact spreads and changes across time is of great interest to scientists and policy-makers in a wide variety of disciplines. If a method for learning this filtering function in a data-driven manner existed, one would be able to extract useful knowledge about the mechanisms that generate variability from a variety of time series data sources.

\begin{figure}[h!]
    \centering
    \includegraphics[width=0.6\textwidth]{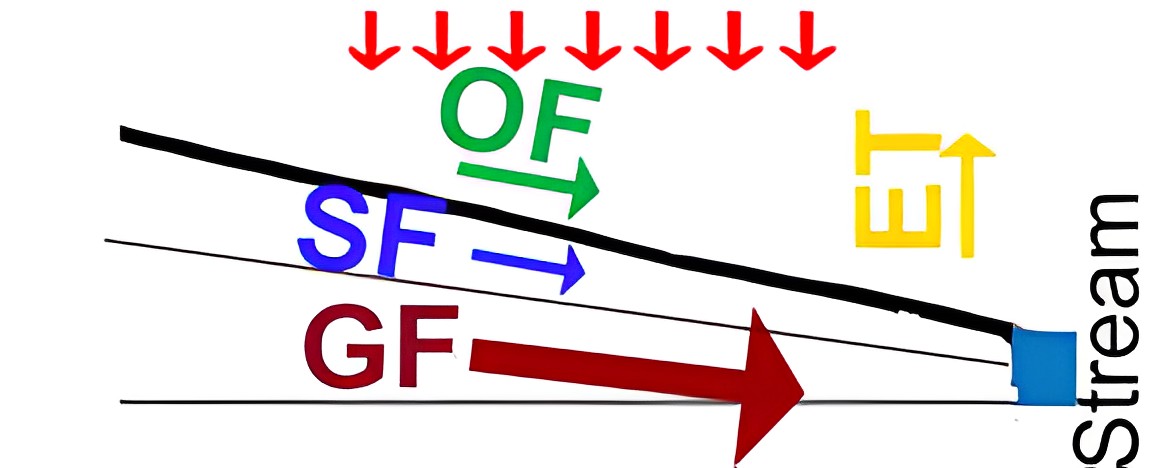}
    \caption{A flow path diagram depicting how precipitation (red arrows) gets partitioned into evapotranspiration (ET), overlandflow (OF), shallow subsurface flow (SF) or deep groundwater flow (GF), with arrow width depicting illustrative relative effect size.}
    \label{fig:flowpath}
\end{figure}

More specifically, this research is motivated by the desire to learn the data generating processes behind streamflow time series. Observational units in hydrology called catchments, filter precipitation into streamflow (Figure \ref{fig:flowpath}). This process may sound simple; however, slightly different geologic, climatic, or topographic features may produce vastly different filtering functions. Understanding the filtering function for a given catchment is vital for a variety of reasons. For example, knowing that a catchment filters most rainfall slowly through subsurface and deep flow pathways would indicate that the stream is vulnerable to soil and deep rock contamination from human activities such as mining \citep{kunz2020towards}. Furthermore, inferring the geological or climatological factors that cause process variation can enable better predictions in regions where streamflow is not measured. Indeed, accurately estimating the functions that filter precipitation into streamflow can help hydrologists understand the physical processes behind complex temporal hydrological data. 

As indicated by \citet{botter2010transport}, for a given catchment, water input ($x$) and streamflow $(y)$ can be causally related via a historical filtering function $(h)$ within a time-varying unit hydrograph model (Equation \ref{eq:tv_unithydro}), where the unit hydrograph is defined as the response of streamflow that appears after a unit pulse of rainfall.

\begin{equation}
    y(t) = \int_{0}^{\infty} h(s,\textbf{P}(\textbf{s}),\textbf{T}(\textbf{s}),\textbf{E}(\textbf{s}),\textbf{L}(\textbf{s})) x(t-s) \, ds
    \label{eq:tv_unithydro}
\end{equation}
Although Equation (\ref{eq:tv_unithydro}) is formulated as a linear convolution between water input ($x$) and the filtering function ($h$), this equation is implicitly nonlinear since the filtering function is changing with time due to changes in antecedent water input (\textbf{P}), evapotranspiration (\textbf{E}), temperature (\textbf{T}), and landscape variables (\textbf{L}). Note that the function $h$ depends on all historical values of its inputs, so the bolded \textbf{s} consists of $t-(s+1),t-(s+2),... 0$. Evaporation or transpiration can alter the effect sizes of the filtering function, since a catchment with high evaporation only has a small portion of water available for filtering into streamflow. On the other hand, the depth to which water flows or gets stored in/on the ground dictates the length of time between a precipitation event and a rise in streamflow \citep{jasechko2016substantial,somers2020review}. Overland flow (OF) is the fastest and shallowest path that precipitation can follow, precipitation filtered into shallow subsurface flow (SF) travels down into the soil and towards the stream at a slower pace, and deep groundwater flow (GF) travels deeply into the ground and is the slowest responding flow path (Figure \ref{fig:flowpath}).

\begin{figure}[h!]
    \centering
    \includegraphics[width=0.494\textwidth]{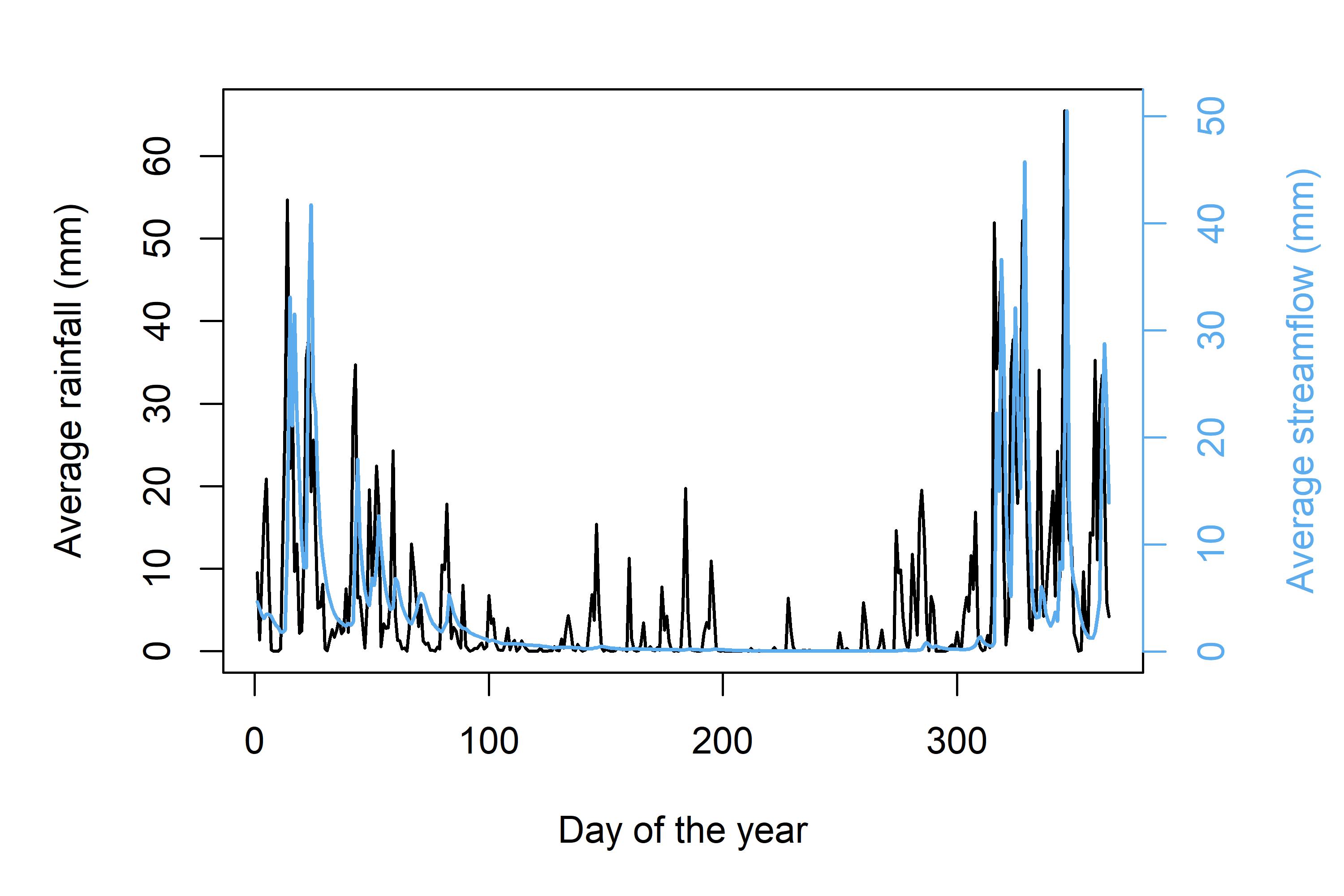} \includegraphics[width=0.494\textwidth]{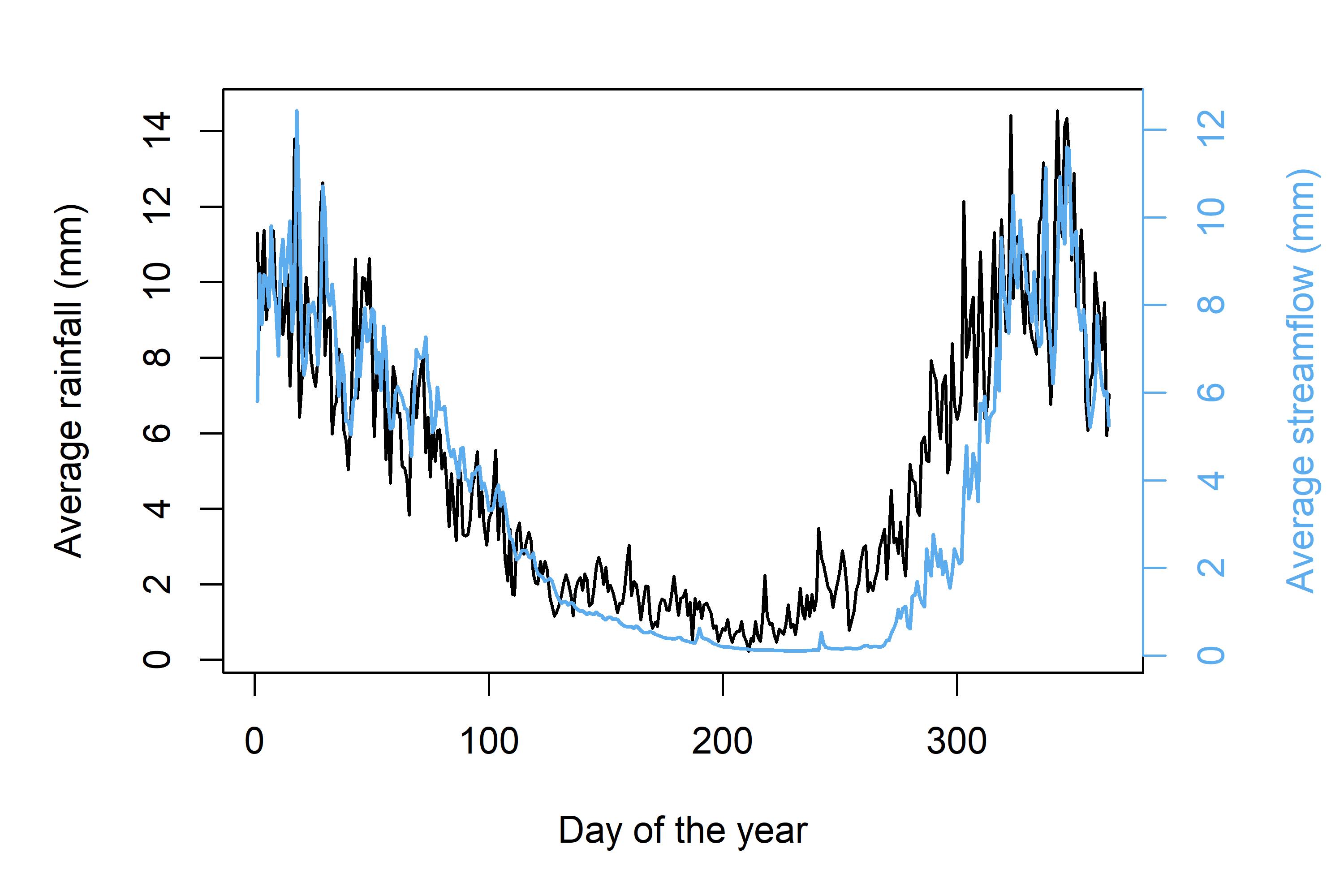} \\
    \includegraphics[width=0.494\textwidth]{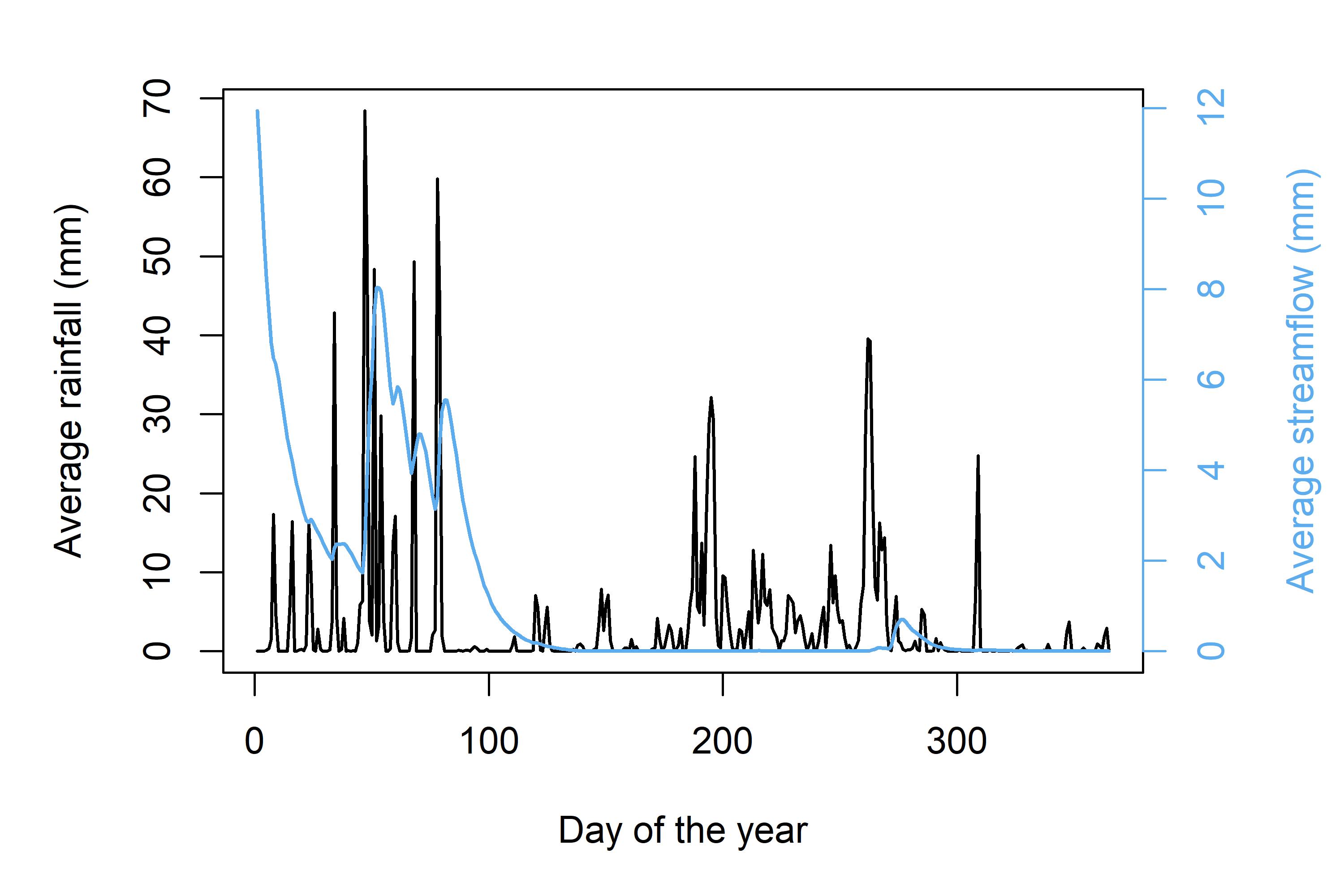} \includegraphics[width=0.494\textwidth]{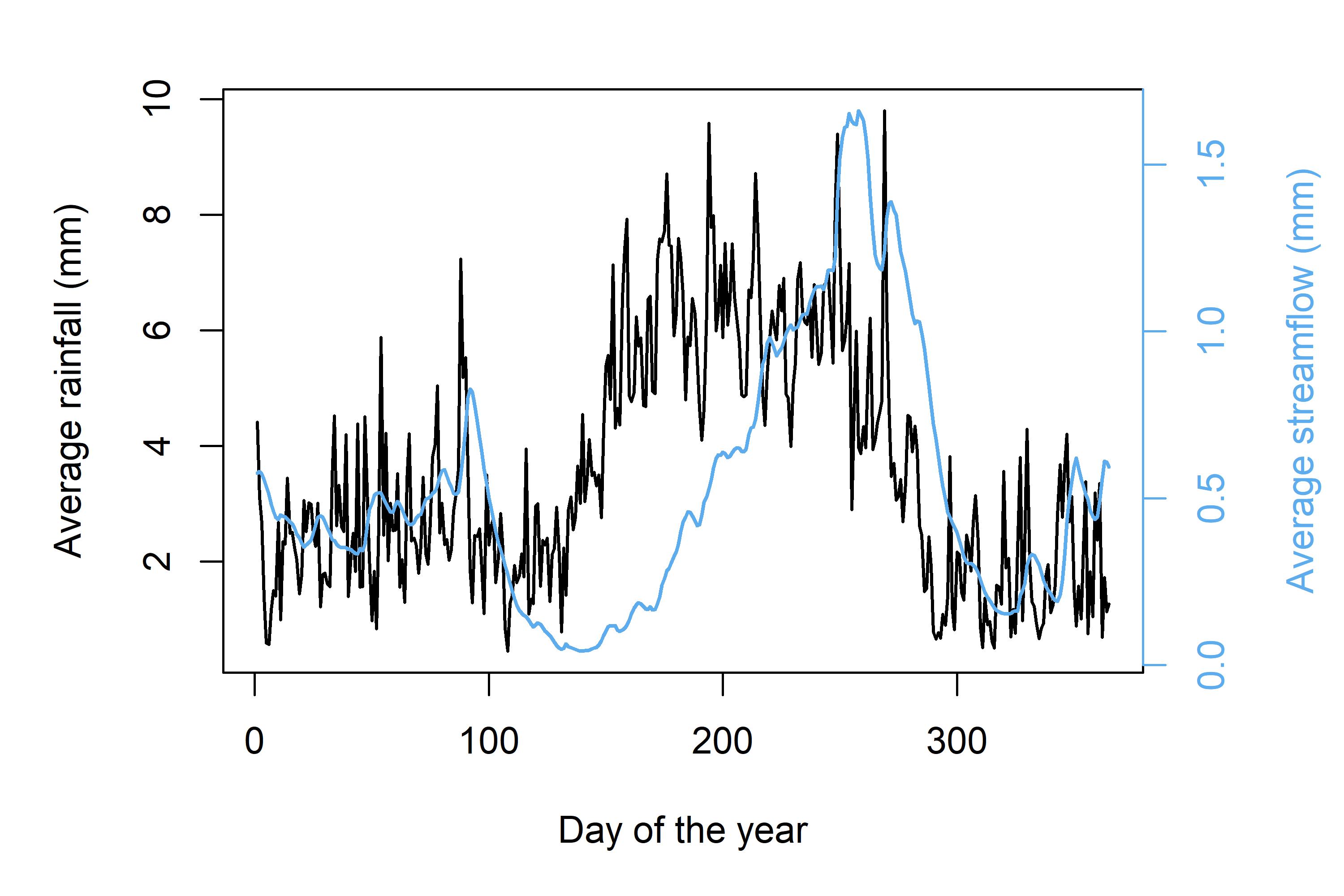} \\
    \caption{Left: The daily streamflow (blue) and rainfall (black) for the Koksilah River (top) and the Withlacoochee River (bottom) for the year 1998. Right: The average daily streamflow (blue) and rainfall (black) for the Koksilah River (top) and the Withlacoochee River (bottom) for 1979-2018.}
    \label{fig:OneYearRainfallStreamflow}
    \label{fig:RainfallStreamflow}
\end{figure}

Streamflow and rainfall time series at the Koksilah River (British Columbia) and the Withlacoochee River (Florida), illuminate the fact that the relationship between rainfall and streamflow is indeed complex and highly nonstationary (Figure \ref{fig:OneYearRainfallStreamflow}). At the Koksilah River, very little winter rainfall is evaporated (days 1-50 and days 320-365), rather it quickly filters through the catchment as streamflow. Conversely, in summer and early fall (days 100-310), any rainfall that may occur does not result in notable streamflow intensification. At the Withlacoochee River, streamflow falls less rapidly during periods of little-to-no rainfall compared to  the Koksilah River (e.g., days 1-50). Evaporation is likely to be high in this catchment, therefore, streamflow may only increase in the event of a large rainfall event as seen on day 50.

The statistical model that most closely follows the causal filtering relation from Equation (\ref{eq:tv_unithydro}) comes from \citet{xun2022sparse} who considered the historical functional linear model,
\begin{equation}
    y_i(t)= \int_{t-\delta}^{t} \beta(s,t) x_i(s) \ ds + \epsilon_i(t).
    \label{eq:historicalLinear0}
\end{equation}
Here, $y_i(t)$ is the outcome of interest at time $t$ of replicate series $i$, $x_i(s)$ is a temporal explanatory variable at the past time $s$ for the same replicate, and $\epsilon_i(t)$ is the residual term. The bivariate coefficient function $\beta(s,t)$ represents the effect of the explanatory variable at past time $s$ on the outcome at the current time $t$, where $t-\delta\leq s \leq t$. To estimate $\beta(s,t)$, \citet{xun2022sparse} minimized an objective function containing three terms: (1) a least-squares loss, (2) a nested group bridge penalty term which reveals the constant lag $\delta$, after which $\beta(s,t)$ is zero, and (3) a smoothness penalty term. This builds upon many previous methods including \citet{malfait2003historical} who originally developed the functional historical linear model, and \citet{harezlak2007penalized} who imposed a discrete difference penalty on the coefficients similar to the P-spline framework \citep{eilers2015twenty}. While both \citet{malfait2003historical} and \citet{harezlak2007penalized} primarily focused on predictive capability, \citet{xun2022sparse} were more interested in accurate parameter estimation. Parameter estimation will also be the focus of this paper. 

The method introduced by \citet{xun2022sparse} is compelling, though it has several notable downsides. First, it is overly complex. Its original formulation is non-convex, necessitating iterative lasso-equivalent optimization steps until convergence. The tent-like basis functions which parameterize $\beta(s,t)$ add computational cost such that the number of basis functions is usually restricted (about 230 basis functions were used in previous works). Second, the triangular finite element framework artificially imposes a constraint on the maximum time lag (i.e., the domain of integration cannot be negative in Equation (\ref{eq:historicalLinear0})). For example, if one were to predict streamflow on day one for the Withlacoochee River in Figure \ref{fig:OneYearRainfallStreamflow}, only one value of rainfall (0mm) could be used, leading to an inaccurate prediction of 1998's peak streamflow. Third, the previous frameworks do not allow for a time dependent time-lag parameter $\delta(t)$. Although it is common to assume an arbitrary fixed window of time for which precipitation can affect streamflow, this is a severe limitation of previous works \citep{janssen2021assessment}, so finding the true underlying $\delta(t)$ is of interest to the hydrological community \citep{tennant2020utility,chiu1969linear}. Other similar methods, with similar issues, can be found in the distributed lag literature \citep{almon1965distributed,pesando1972seasonal,rushworth2013distributed,asencio2014functional,gupta1968efficient,liao2023analysis}.

% In our motivating problem, streamflow on some day of the year $t$ is a function of $\delta(t)$ days of historical rainfall. Both streamflow and rainfall are measured as daily accumulations in millimeters, so the data is equally spaced. Ideally, we should be able to learn $\delta(t)$ and the function $\beta(s,t)$ that different catchments employ to turn rainfall into streamflow at a daily scale, even in the presence of substantial noise. Given these circumstances which are not only found in hydrology but also across all areas of Earth sciences, \citet{eilers2003perfect} suggests using the identity matrix as a basis which is equivalent to zero-degree P-splines. This method is referred to as Whittaker smoothing \citep{whittaker1922new}, and it vastly simplifies the degree-three tensor product P-splines used by \citet{rushworth2013distributed} and the tent function finite element framework employed by \citet{malfait2003historical}, \citet{harezlak2007penalized}, and \citet{xun2022sparse}, while not hurting performance and improving speed \citep{eilers2003perfect,eilers2015twenty,eilers2021practical}. This avoids the need to perform costly preprocessing, removes the need to tune the number of splines \citep{eilers2021practical}, and removes the need to assume different functional replicates are independent.

This article has three major contributions. First, to the best of our knowledge, this is the first attempt to estimate a time-dependent time-lag parameter in historical functional linear models. Second, we simplify and increase the flexibility of previous historical functional linear model methods by replacing the tent-like basis functions with the tensor product Whittaker basis such that higher resolution coefficient and time-lag functions can be recovered. Finally, we successfully demonstrate that our method can lead to novel research directions within hydrology by accurately estimating the ground-truth filtering function $\beta(s,t)$ and time-dependent time-lag parameter $\delta(t)$.

% The remainder of this paper is organized as follows. Our method is formally presented in Section \ref{sec:Methods} along with a sketch of its algorithm. In Section \ref{sec:experiments}, we conduct several experiments on real streamflow data and compare the results to expert knowledge.  Next, eight simulation experiments in Section \ref{sec:simStudies} assess the quality of parameter estimation at different levels of autocorrelated noise. Finally, in Section \ref{sec:conclustion}, we summarize the work, explain the limitations of our method and study, and discuss ideas for future work.

\section{Historical functional linear model with dynamic sparsity (HFLM-DS)}
\label{sec:Methods}

Let $y_i(t)$ be the observed response at time $t \in 1 \dots T$ for replicate $i \in 1...n$, and let $x_i(t)$ be the corresponding covariate. In our application, $y_i(t)$ is daily streamflow during the $i$-th year, and $x_i(t)$ is daily rainfall. We begin by considering the exact causal streamflow relation described in Equation (\ref{eq:tv_unithydro}), where \textbf{V} is the set of all important antecedent variables \{\textbf{P},\textbf{T},\textbf{E},\textbf{L}\}. We then separate the true filtering function $h(s,\textbf{V})$ into an identifiable component that is stable across years due to consistent seasonal variations of antecedent conditions ($\beta(s,t)$) and an unidentifiable component that changes from year to year ($h'(s,\textbf{V})$).

\begin{align*}
        y_i(t) & = \int_{0}^{\infty} h(s,\textbf{V}) x_i(t-s) \ ds & &\textnormal{Exact filtering function model}\\
        & = \int_{0}^{\infty} \left[ \beta(s,t) + h'(s,\textbf{V}) \right] x_i(t-s) \ ds & &\textnormal{Expand $h$ into subcomponents}\\
        & = \int_{0}^{\infty} \beta(s,t) x_i(t-s) \ ds + \int_{0}^{\infty} h'(s,\textbf{V}) x_i(t-s) \ ds & &\textnormal{Distribute and separate terms}\\
        & = \int_{0}^{\infty} \beta(s,t) x_i(t-s) \ ds + \epsilon_i(t) & & \textnormal{Unparamaterized term as error}
    \end{align*}

\begin{figure}[h!]
    \centering
    \includegraphics[width=0.75\textwidth]{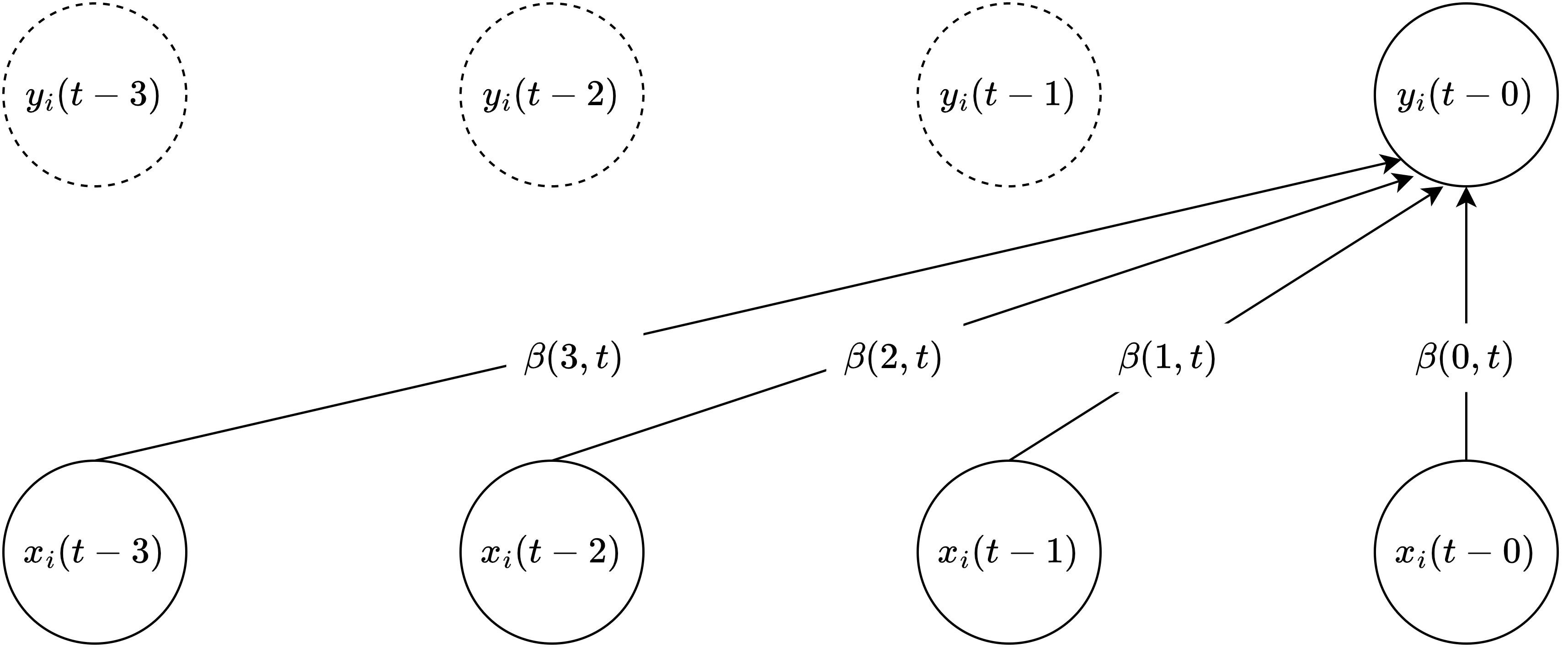}
    \caption{The relationships between the response $y$ and the cause $x$ is shown for D=4.}
    \label{fig:CausalGraph}
\end{figure}

After separating terms, letting the unidentifiable part become the autocorrelated error term $\epsilon_i(t)$, and letting $D$ be the total number of considered lags, our functional historical linear model is given by 
\begin{equation}
    y_i(t)=  \int_{0}^{D-1} \beta(s,t) x_i(t-s) \ ds + \epsilon_i(t),
    \label{eq:historicalLinear}
\end{equation} 
and the systematic part is visualized in Figure \ref{fig:CausalGraph}.

As with \citet{xun2022sparse} we assume without loss of generality that the response and covariate have had their seasonal signals removed such that they have mean zero for each day of the year. Not only does the usage of rainfall and streamflow anomalies remove the need for a functional intercept term, it also removes any potential periodic confounding effects present in both signals \citep{moges2022strength}. We note that $t-s$ can be negative. When this occurs, we draw from the end of the previous replicate $i-1$. 

% In the same way as \citet{rushworth2013distributed}, we treat the data as time series since the data is equally-spaced, and smoothing via functional data analysis as was done in \citet{xun2022sparse} and \citet{harezlak2007penalized} can remove extreme points which are of great interest to Earth scientists. Further, both streamflow and rainfall are non-smooth (Figure \ref{fig:OneYearRainfallStreamflow}), so representing them via smooth functions is not recommended. 

Let $\phi_1(s,t),..., \phi_K(s,t)$ denote a sequence of $K$ known basis functions. Given these basis functions and their corresponding basis coefficients $\textbf{b}= [b_1,..., b_K]^T$, the coefficient function $\beta(s,t)$ can be denoted as

\begin{equation*}
    \beta(s,t) = \sum_{k=1}^K b_k \phi_k(s,t).
\end{equation*}
Using this basis expansion, we may reformulate Equation (\ref{eq:historicalLinear}) as

\begin{eqnarray*}
    y_i(t) &=&  \int_{0}^{D-1} \sum_{k=1}^K b_k \phi_k(s,t) x_i(t-s) \ ds + \epsilon_i(t) \\
    &=& \sum_{k=1}^K b_k \int_{0}^{D-1} \phi_k(s,t) x_i(t-s) \ ds + \epsilon_i(t)\\
     &=& \sum_{k=1}^K b_k z_{ik}(t) + \epsilon_i(t),
    \label{eq:historicalLinearSimp}
\end{eqnarray*}

where $$z_{ik}(t)= \int_{0}^{D-1} \phi_k(s,t) x_i(t-s) \ ds.$$

As stated in the introduction, one of our foremost goals is to infer the dynamic time lag $\delta(t)$ after which the cause has no discernible effect on the response, i.e.,
\begin{equation}
    \begin{split}
     \delta(t)=  \max \{s\} \hspace{10mm}
    \text{subject to } \beta(s,t) \neq 0.
    \label{eq:deltat}
\end{split}
\end{equation}
The accuracy in determining $\delta(t)$ is directly related to the resolution of the basis functions that parameterize $\beta(s,t)$, but increasing the number of bases can be computationally prohibitive if we continue the use of the tent-like basis functions utilized in previous works \citep{xun2022sparse}. Under these circumstances, using the Whittaker basis is the most appropriate solution \citep{eilers2003perfect,whittaker1922new}. In this work, we parameterize the coefficient function $\beta(s,t)$ with tensor products of Whittaker basis functions $\phi_k(s,t)=w_m(s) \otimes w_{\ell}(t)$ for $m \in \{0,..., D-1\}$ and $\ell \in \{1,..., T\}$, where $m=k - 1 - (\lceil k/D \rceil -1) \times D$ and $\ell= \lceil k/D \rceil$. Here $w_m(s)$ and $w_{\ell}(t)$ denote the Whittaker basis functions, which are equivalent to zero-degree B-spline basis functions with a knot spacing of one \citep{eilers2003perfect}:

$$
    w_p(s)= \begin{cases} 
      1 & p \leq s < p+1 \\
      0 & \texttt{otherwise}
   \end{cases}.
$$

Suppose the maximum reasonable lagged influence between $x_i(t)$ and $y_i(t)$ for all $t$ occurs at lag $s=D-1$, then we must estimate $K = DT$ parameters $\{b_k:k= 1,...,K\}$. In general we expect that $nT = N \ll K$, thus the system is often underdetermined. This problem cannot be overcome without some assumptions about the structure of the coefficient function $\beta(s,t)$, or equivalently the coefficients $b_k$, and we therefore stipulate three major assumptions. First, for the same lag, the coefficients at consecutive time points should have similar coefficient values (e.g., $\beta(s,t) \approx \beta(s,t+1)$). Second, for the same time point, coefficients at consecutive lags should have similar values (e.g., $\beta(s,t) \approx \beta(s+1,t)$). Finally, we expect $\beta(s,t)$ to be fairly sparse. Before fitting a model, the extent of the sparsity is unknown, though we expect the sparsity will be in regions with larger lags. We limit the number of assumptions so that our methods are flexible and widely applicable \citep{clark2011pursuing}, though we could have further assumed that the coefficient function should be non-negative for our application since additional precipitation cannot reduce streamflow. Instead of adding this constraint, we keep our model flexible to allow an additional evaluation of the scientific validity of our method.

\begin{figure}[h!]
    \centering
    \includegraphics[width=0.5\textwidth]{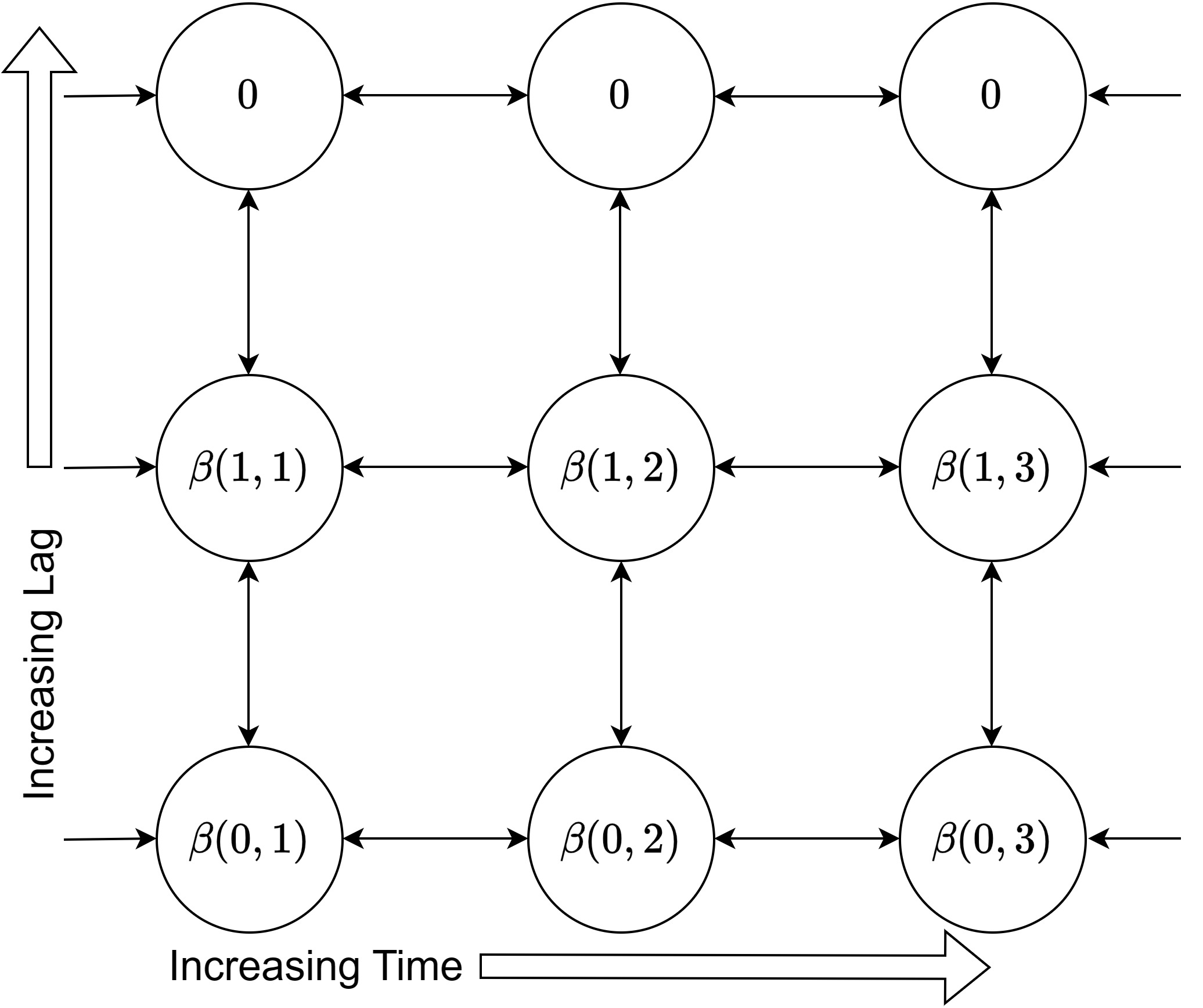}
    \caption{The general structure of $\beta(s,t)$ is shown for $D=2$ and $T=3$, where $s$ is lag and $t$ is time. The bidirectional lines between nodes indicate that the nodes should have similar values. The unidirectional lines indicate that there is periodic behaviour with a period of $T+1$ (i.e., $\beta(0,3)$ and $\beta(0,1)$ are connected). The bottom row of nodes represents the coefficients relating the current time's cause to the current time's response while the next rows represent the coefficients for consecutive past lags. The top row shows the zero boundary conditions.}
    \label{fig:modelDiag}
\end{figure}

With these three assumptions in mind, we now impose regularization and sparsity constraints on $\beta(s,t)$. With the first assumption, we regularize the squared differences of the coefficient function at consecutive points in time. This can be done with the horizontal first difference penalty matrix ($D_H$), which was first applied to functional historical linear models in \citet{harezlak2007penalized} and further used by \citet{xun2022sparse}. We further add periodic boundary conditions to $D_H$ in a similar fashion as \citet{garcia2010robust}, such that $\beta(s,T) \approx \beta(s,1)$. Likewise, given the second assumption, we penalize the squared differences between coefficients with the same time but with consecutive lag values by introducing the vertical first difference penalty matrix ($D_V$). We additionally consider a zero top boundary condition on the vertical penalties since lags past $D-1$ are assumed to have zero influence. The structure of the coefficients and the penalties are visualized in Figure~\ref{fig:modelDiag}. Further, if we take $D=2$, $T=3$, and $K=6$, then $D_H$, $D_V$, and $\beta(s,t)$ are given by:
\begin{align*}
    D_H = \begin{bmatrix} 
    -1 & 0 & 1 & 0 & 0 & 0 \\
    0 & -1 & 0 & 1 & 0 & 0\\
   0 & 0 & -1 & 0 & 1 & 0\\
   0 & 0 & 0 & -1 & 0 & 1 \\
   1 & 0 & 0 & 0 & -1 & 0 \\
   0 & 1 & 0 & 0 & 0 & -1 \\
\end{bmatrix} \hspace{.05cm} D_V = \begin{bmatrix} 
    -1 & 1 & 0 & 0 & 0 & 0 \\
    0 & 0 & -1 & 1 & 0 & 0\\
   0 & 0 & 0 & 0 & -1 & 1\\
   0 & 1 & 0 & 0 & 0 & 0 \\
   0 & 0 & 0 & 1 & 0 & 0 \\
   0 & 0 & 0 & 0 & 0 & 1 \\
\end{bmatrix} \hspace{.05cm}  \beta(s,t) = \begin{bmatrix} 
    \beta(0,1) \\
    \beta(1,1)\\
    \beta(0,2)\\
    \beta(1,2) \\
    \beta(0,3) \\
    \beta(1,3)
\end{bmatrix}.
\end{align*}

% In statistics, sparsity is usually induced via the lasso penalty \citep{tibshirani1996regression}, however, it does not have the oracle and sign consistency properties, meaning it cannot distinguish which coefficients should be exactly zero according to the data-generating process. This was shown experimentally in the original lasso paper \citep{tibshirani1996regression}, and it was shown theoretically in \citet{meinshausen2009lasso}, \citet{zhao2006model}, \citet{fan2001variable}, and \citet{leng2006note}. The oracle and consistency properties of bridge penalties are better than lasso, though they still fail when the number of parameters exceeds the number of observations and when there is extensive multicollinearity \citep{huang2008asymptotic}. Both of these troublesome properties are expected in problems surrounding our intended application. As an alternative, \citet{zhao2006model} suggests using the L0 penalty, though this poses a new problem of non-convexity and NP-hard computational complexity \citep{huo2007stepwise}. The L0-norm has the oracle property and is selection consistent, meaning it will select the correct features which are non-zero in the ground-truth model with a probability converging to 1 under weaker conditions compared to the lasso \citep{zhang2012general,staerk2018adaptive}. Due to the strongly hierarchical nature of our model, optimizing the L0-norm problem changes from $O(2^K)$ to $O(K)$, thus, for our application it is computationally tractable and theoretically preferable for variable selection.

Most previous works that have developed historical functional linear models have imposed sparsity in their model \citep{harezlak2007penalized,malfait2003historical,xun2022sparse}, however, they all assume that the lag after which the explanatory feature has no effect on the response is static in time. This assumption is indeed too strong for many applications, especially in Earth science. Instead, we impose sparsity with a more flexible approach via nested group 2-norm thresholding such that a dynamic time lag $\delta(t)$ can be reliably estimated. The nested group framework allows us to place each coefficient in a series of groups according to a hierarchical structure \citep{zhao2009composite}. Let $A_{0,t}, A_{1,t}, ..., A_{D-1,t}$ be a series of groups defined for each time point $t \in 1,..., T$. The group $A_{s,t}$ contains the indices of the coefficients corresponding to $\beta(s,t), ..., \beta(D-1,t)$. Notice that the groups are nested such that $ A_{D-1,t} \subset A_{D-2,t} \subset ... \subset A_{0,t}$. These groups will be further discussed when our method and algorithm are detailed in Section \ref{sec:algo}.

\subsection{Autocorrelated errors}

As we illustrated in the derivation at the beginning of this section, our model is somewhat misspecified due to replacing dynamic components such as antecedent temperature with the day of the year $t$ \citep{granger1974spurious}. While $t$ serves as a good proxy for these dynamic antecedent conditions, it is imperfect due to climatic variations across various years. This imperfection as well as inherent day-to-day autocorrelation in streamflow and possible systematic measurement errors \citep{horner2018impact,sorooshian1980stochastic,schoups2010formal,kim2023time}, can lead to strongly autocorrelated errors \citep{sun2021adjusting,thursby1987ols}. For example, suppose that for a given catchment, May is a dry month across most years, but for a particular year, it is extremely wet. Our model will likely choose $\beta(s,t)$ to be fairly small in May to fit the data across most years, however, for the particularly wet May, our model will systematically under predict streamflow (i.e., autocorrelated errors) since streamflow has persistent magnitudes and our $\hat{\beta}(s,t)$ is too low for May in that particular year. If we ignore this property of the errors, the coefficient estimates will remain unbiased, but the estimates will be inefficient with high variance, likely leading to poor overall recovery of the underlying causal process \citep{hibbs1973problems}. Naturally, one could consider previous lags of the response to improve predictive capabilities and reduce autocorrelation, as is done in the classical time series framework ARIMAX \citep{box2015time}, however, prioritizing forecasting accuracy leads to several issues. First, previous streamflow does not cause future streamflow, thus incorporating it into the model leads us further from the true data-generating processes thereby instigating incorrect causal effect estimates \citep{dafoe2018nonparametric,barnett2017time}. Further, while data for precipitation is globally available and accurate, other data such as streamflow or evaporation is usually inaccurate or unavailable. Since hydrologists often require a model that is easily transferable to all ungauged or data-limited regions \citep{janssen2021hydrologic}, explicit use of previous streamflow values must be excluded from the model.

The most common way to account for autocorrelated errors while not explicitly using lagged response values in the model, has been the Cochrane-Orcutt or Prais-Winsten correction method \citep{cochrane1949application,prais1954trend}. Unfortunately, both Cochrane-Orcutt and Prais-Winsten corrections failed to outperform our method without autocorrelation correction for parameter estimation in our initial simulation studies. It is unclear exactly why these methods failed, but we have several suspicions. First, we noticed that artificially reducing the smoothness penalty term found from our validation set led to better results, but there was no consistent reduction we could apply to produce optimal results. Second, we noticed that if we only consider a few lags (i.e., $D<5$), the corrective methods far outperformed default regression, however if $D>100$, default regression far outperformed the corrective methods. Reviewing the causes of these and other issues could be an interesting direction for future work \citep{dagenais1994parameter,sims1972role,mcguirk2009revisiting,mizon1995simple,dafoe2018nonparametric,thursby1987ols}.

%Both of these methods rely on an initial estimation of the model parameters assuming uncorrelated errors, followed by fitting an AR(p) model to the residuals. Using the AR(p) terms, the input and output time series are quasi-differenced and the model is refit for the final parameter estimates. Both sets of estimators are asymptotically efficient, but the Prais-Winsten procedure can be advantageous due to saving the first observation, especially in low sample size settings \citep{dielman1989small,doran1981omission,beach1978full}.

Instead, we found that using an autoregressive distributed lag model followed by deconvolution achieved better results in the presence of autocorrelated errors. This method is comprehensively explained in \citet{kirchner2022impulse} but also appears in many earlier works \citep{baltagi2011distributed,pagano1981fitting,tsay1985model,young2002advances}. Briefly, the equations
\begin{equation*}
    y_t = \alpha_1 y_{t-1} +b_0 x_t + b_1 x_{t-1} + b_2 x_{t-2} + \epsilon_t
\end{equation*}
\begin{equation*}
    y_t = \beta_0 x_t + \beta_1 x_{t-1} + u_t
\end{equation*}
are equivalent since we can subtract the second equation by its lagged self multiplied by $\alpha_1$ to obtain
\begin{equation*}
    y_t = \alpha_1 y_{t-1} +\beta_0 x_t + (\beta_1 - \alpha_1 \beta_0) x_{t-1} - \alpha_1 \beta_1 x_{t-2} + \epsilon_t,
\end{equation*}
when $u(t)$ is an AR(1) process defined by $u_t=\alpha_1 u_{t-1} + \epsilon_t$. Clearly, we can simply solve the first equation without worry of autocorrelated errors, then use deconvolution to obtain the $\beta$ coefficients of interest from the modified $b$ coefficients (see \citet{kirchner2022impulse} for further details). The above simple example is easily applied to our more complex methodology and implemented in the algorithm below. Fortunately, this also reduces the computational cost compared to the Prais-Winsten procedure due to only requiring a single model fit, with a diagnostic test of no autocorrelation, followed by a simple deconvolution instead of two model fits. Note that all autocorrelated error correction approaches implicitly assume the common factor restriction implying that the response does not cause the lagged explanatory variables \citep{mcguirk2009revisiting}, a safe assumption for our application.

\subsection{Algorithm}
\label{sec:algo}

Let the response $\textbf{Y}$ be defined as a column vector of length $N-(D-1)$. We then define the sparse matrix $\textbf{Z} \in \mathbb{R}^{N-(D-1) \times K}$ which has $D$ non-zero entries in each row corresponding to the valid entries of $\beta(s,t)$ given the current time of $y_i(t)$. 
Here, $\textbf{b}$ is the vector of $b_k \ \forall k \in 1,...,K$. For example, suppose $n=2$, $D=2$, $T=3$, and $K=6$, then

\begin{align*}
    \textbf{Y} = \begin{bmatrix} 
    y_1(2) \\
    y_1(3)\\
   y_2(1)\\
   y_2(2) \\
   y_2(3)
\end{bmatrix} \hspace{0.5cm} \textbf{Z} = \begin{bmatrix} 
    \cancelto{0}{z_{1,1}(2)} & \cancelto{0}{z_{1,2}(2)} & z_{1,3}(2) & z_{1,4}(2) & \cancelto{0}{z_{1,5}(2)} & \cancelto{0}{z_{1,6}(2)} \\
    \cancelto{0}{z_{1,1}(3)} & \cancelto{0}{z_{1,2}(3)} & \cancelto{0}{z_{1,3}(3)} & \cancelto{0}{z_{1,4}(3)} & z_{1,5}(3) & z_{1,6}(3)\\
   z_{2,1}(1) & z_{2,2}(1) & \cancelto{0}{z_{2,3}(1)} & \cancelto{0}{z_{2,4}(1)} & \cancelto{0}{z_{2,5}(1)} & \cancelto{0}{z_{2,6}(1)}\\
   \cancelto{0}{z_{2,1}(2)} & \cancelto{0}{z_{2,2}(2)} & z_{2,3}(2) & z_{2,4}(2) & \cancelto{0}{z_{2,5}(2)} & \cancelto{0}{z_{2,6}(2)} \\
   \cancelto{0}{z_{2,1}(3)} & \cancelto{0}{z_{2,2}(3)} & \cancelto{0}{z_{2,3}(3)} & \cancelto{0}{z_{2,4}(3)} & z_{2,5}(3) & z_{2,6}(3)
\end{bmatrix} \hspace{0.5cm} \textbf{b} =\begin{bmatrix} 
    b_1 \\
    b_2\\
    b_3\\
    b_4 \\
    b_5 \\
    b_6
\end{bmatrix}.
\end{align*}

Our method first estimates a smooth coefficient function via:

\begin{equation}
    \label{eq:OurSmoothOpt}
    \min_{\textbf{b}}||\textbf{Y}-  \textbf{Z}\textbf{b}||_2^2 + w_h ||D_H \textbf{b}||_2^2 + w_v ||D_V \textbf{b}||_2^2,
\end{equation}

and the group 2-norms are computed using this initial coefficient function estimation (see the end of Section \ref{sec:Methods}). Then, a sequence of models are fit using coefficients with corresponding group norms greater than a sequence of $q$ values in
\begin{equation*}
    %\label{eq:OurSparseOpt}
    \min_{\textbf{b}}||\textbf{Y}- \textbf{Z} \textbf{b}||_2^2 + w_h ||D_H \textbf{b}||_2^2 + w_v ||D_V \textbf{b}||_2^2 \text{   s.t. } \textbf{b}_{A_{s,t}}=0 \text{ when } ||\textbf{b}_{A_{s,t}}||_2^2<q.
\end{equation*}

Once the optimal $q$ is chosen (selected via a user's choice of methods), previous response values are added as covariates and modified coefficients are computed. Finally, the coefficients are deconvolved back into their intended form. A detailed description of our algorithm is provided in Algorithm \ref{algo:ourAlgo}.

\begin{algorithm}[h!]
\caption{}\label{algo:ourAlgo}
\begin{algorithmic}[1] 
\REQUIRE{One response time series $Y=[y_1, ..., y_N] $ and one explanatory time series $X=[x_1,..., x_N]$}. \\
\STATE Find optimal weights $w_h,w_v$ such that some objective function (i.e, $R^2$) is maximized on a validation set.
\STATE Find optimal $\textbf{b}$ using the optimal weights $w_h$ and $w_v$ and the least squares criterion in Equation (\ref{eq:OurSmoothOpt}). \\
\STATE Compute the group norms $||\textbf{b}_{A_{s,t}}||^2_2$ for all $(s,t)$.
\STATE Find optimal sparsity threshold $q$ to obtain $\delta(t)$.
\STATE Given $q$, recompute the optimal smoothing weights, $w_h$ and $w_v$, and refit the model using all data and $y_{t-1}... y_{t-c}$ as additional covariates.
\STATE Deconvolve the optimized coefficients to obtain $\beta(s,t)$.
\RETURN $\beta(s,t)$ and $\delta(t)$
\end{algorithmic}
\end{algorithm}

\subsection{Confidence intervals}

In order to approximate our uncertainty about our estimated $\beta(s,t)$ and $\delta(t)$ parameters of interest, we can construct bootstrap confidence intervals. However, ordinarily, the use of bootstrap to construct confidence intervals is strongly dependent on the assumption of independent observations \citep{kunsch1989jackknife}. To rectify this issue, we suggest using the block residual bootstrap as described in \citet{asencio2014functional} or \citet{paparoditis2003residual}. In essence, we fit our model without corrections for autocorrelated errors and estimate the residual series, then we can divide the residual series into blocks with lengths of a single year. Finally, bootstrap samples can be formed by sampling blocks with replacement and adding them to the series of model predictions. We do not view confidence intervals as directly important for our application, so we do not pursue this method during our experiments, but residual and block bootstraps have been extensively tested in previous works \citep{chatterjee2011bootstrapping,lahiri1999theoretical,paparoditis2003residual}.

\section{Hydrology Data Application}
\label{sec:experiments}

We apply our methodology to data from two diverse catchments to show its versatility for hydrology. For the real data experiments, the relationship between actual rainfall and streamflow is modelled. The simulation study also uses real rainfall data, to mimic actual patterns in the time series, but streamflow is simulated according to known coefficient functions with known lag structures, so we can assess estimation performance.

The first catchment we explore is the Koksilah River which is located in Cowichan, British Columbia, Canada. The area of this catchment is $236\text{km}^2$ with an average elevation of 461 meters. About 11\% of precipitation falls as snow and it is an extremely wet catchment with an aridity index (fraction of potential evapotranspiration over precipitation) of just 0.37. The second catchment we explore is the Withlachoochee River which is located in Dade City, Florida, United States. The area of the catchment is $650 \text{km}^2$, and it is low lying with an average elevation of only 42 meters above sea level. It does not experience any snow and it is a fairly dry catchment with an aridity index of 1.07. To obtain rainfall for both these areas, we use EMDNA \citep{tang2021emdna}, a high-quality climate dataset that has complete daily precipitation and temperature values for 1979-2018 at 10km grid squares across North America. From this data, daily rainfall values are generated for the Koksilah and Withlacoochee rivers by averaging precipitation and temperature values across the catchments, then using a temperature threshold of $0 ^{\circ}C$, we compute daily rainfall (Figure \ref{fig:RainfallStreamflow}). Leap days are removed, leaving us with $365*40 = 14,600$ observations. We decided that the maximum reasonable lag should be $D=150$ days. Indeed, regardless of the catchment location, all precipitation in rain-dominated catchments should drain, become captured by deep groundwater systems, or be evaporated from the catchment within 150 days (5 months) \citep{jasechko2016substantial,tennant2020utility}, leaving a total of $365 \times 150= 54,750$ parameters to be estimated to construct $\beta(s,t)$ (Equation \ref{eq:OurSmoothOpt}). The choice of maximum lag also alters the total number of available observations since to predict the response we require the previous 150 days of explanatory data. Therefore, the final number of observations is 14,451.

Hyperparameters $w_h$ and $w_v$ are optimized with Bayesian optimization using the \textit{GPfit} R package \citep{macdonald2015gpfit}. We start with 30 uniformly random initial points, then continue by choosing the next point by maximizing the expected increase in validation set $R^2$ for an additional 35 iterations. After experimenting on several catchments, we noticed that the optimal $w_h$ is often much larger than the optimal $w_v$, thus we set our search range as $e^{[8,24]}$ for $w_h$ and $e^{[-5,15]}$ for $w_v$.

Choosing the optimal threshold $q$ is often the most difficult step in Algorithm \ref{algo:ourAlgo}. We do not specifically define how this should be done in the methods section as we leave it up to the user and their goals, though we will give some guidance in the context of our hydrological application. If the goal is feature selection, where removing a few important features is not as critical as only selecting features which are surely important, greater sparsity may be desired. If instead, a lag should only be removed if it is surely unimportant, such as in sure independence screening \citep{fan2008sure}, future users may desire less sparsity. Minimizing errors on the validation set seemed to give high test set $R^2$ scores, however, the final coefficient function indicated too little sparsity. The phenomena was also found in \citet{rushworth2013distributed} who observed optimizing the AIC leads to too little smoothness, pushing the authors to use the AIC-optimal smoothness parameter as a lower bound and artificially increasing this value. In all experiments below, we find the "knee-point" as outlined in \citet{satopaa2011finding} to choose $q$ since it gave consistently strong results. 

% A popular alternative method to this is choosing the "knee point" in the plot of $q$ versus the whole dataset $R^2$ \citep{satopaa2011finding}. This method is popular for unsupervised learning such as clustering analysis, though for our purposes, it leads to too much sparsity. Instead of occurring at the middle of the knee, the ground-truth sparsity threshold seemed to often occur at the beginning of the knee, i.e., "knee-onset" \citep{fermin2020identification}. One promising way to automatically find the "knee-onset" is maximizing Menger curvature between the first, last, and current point \citep{satopaa2011finding}, though we found that this method was too sensitive to noise. 

We let $c=2$ for all experiments, meaning in step 5 of Algorithm \ref{algo:ourAlgo}, we add the previous two values of the response as additional covariates in our autoregressive distributed lag model (i.e., the errors are modelled as an AR(2) process). This is a conservative decision since we will also be able to accurately model AR(1) errors \citep{schmidt1971estimation}. All experiments were run in R version 4.2.1.

\subsection{Application to actual streamflow}
\label{sec:hydroStudy}

Daily streamflow data from the Koksilah River and the Withlachoochee River were gathered from the Environment Canada HYDAT database and from CAMELS \citep{addor2017camels}, respectively. The average daily streamflow for both rivers is visualized in Figure~\ref{fig:RainfallStreamflow}. To improve predictive performance and remove patterns and non-Gaussianity from the residual plots, we transform streamflow $y$ via $y_{new} = \log(y+1)$. The estimates of our hyperparameters $w_h$ and $w_v$ are obtained after splitting the data with 80\% of the data for training and 20\% of the data for validation. The final estimates of $\beta(s,t)$, $\delta(t)$, and the whole dataset $R^2$ are then obtained by training on all available data. We also evaluate the predictive performance of our model with a test set. To compute test set $R^2$ values, we split the data with 60\% for training, 20\% for optimizing the hyperparameters $w_h$ and $w_v$, and 20\% for testing.

\begin{figure}[h!]
    \centering
     \begin{subfigure}[b]{0.494\textwidth}
         \centering
         \includegraphics[width=\textwidth]{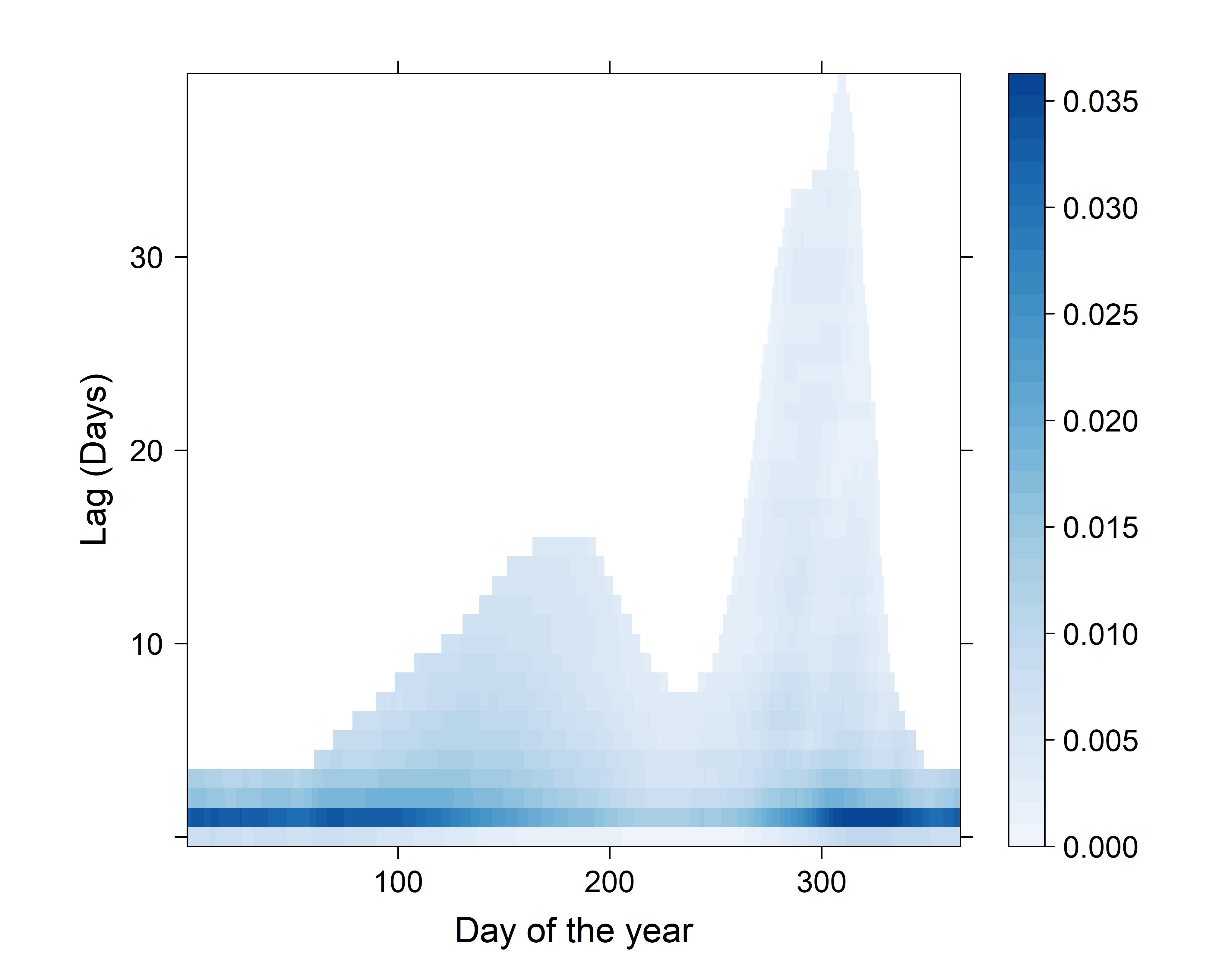}
         \caption{Koksilah River}
         \label{fig:koksilahResults}
     \end{subfigure}
     \begin{subfigure}[b]{0.494\textwidth}
         \centering
         \includegraphics[width=\textwidth]{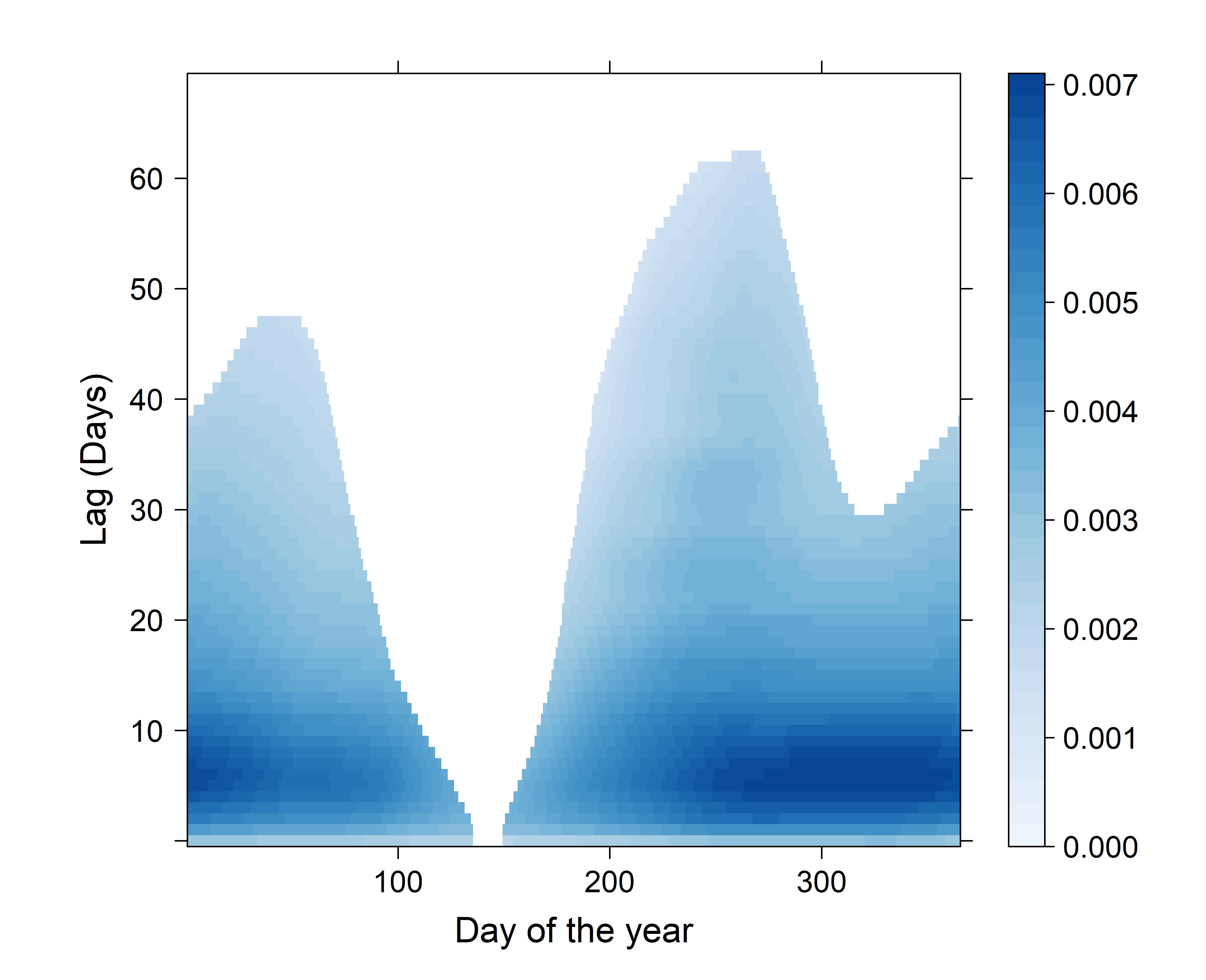}
         \caption{Withlachoochee River}
         \label{fig:WithlachoocheeResults}
     \end{subfigure}
    
    \label{fig:wi}

    \caption{The estimated function $\hat{\beta}(s,t)$ for each lag $s$ and time $t$ for the Koksilah River located in British Columbia (left) and Withlachoochee River, located in Florida (right) from the whole data period (1979-2018).}
\end{figure}

The estimates $\hat{\beta}(s,t)$ for the Koksilah River after training on the entire dataset are shown in Figure \ref{fig:koksilahResults}. In the final model, the autocorrelated errors were modeled via an AR(2) process with fitted coefficients (0.704, 0.122), showing substantial autocorrelation. In winter, rainfall frequency and magnitude peaks after a rapid ramp up from September to October to November (days 260-320) (Figure \ref{fig:RainfallStreamflow}), therefore, by December (day 335), the soil in the catchment is fully saturated and groundwater is fully connected. When this happens, any additional rain either immediately transmits to flood waters heading towards the stream, or the additional rainfall quickly pushes the water already in the soil towards the stream. Further, this behavior can be seen in Figure \ref{fig:RainfallStreamflow}, where December, January, and February (days 330--70) streamflow correlates strongly with rain falling the previous day. Therefore, observing a $\delta(t)$ of around 3 days from day 350--60 empirically confirms and supports existing expert knowledge. Two distinct peaks of maximum lag appear in summer and fall in Figure \ref{fig:koksilahResults}. The peak in summer at around day 190, with a $\delta(t)$ of about 18 days displays a long term recession behavior in the catchment. Starting in March, this catchment appears to enter a drying phase where evapotranspiration and runoff exceed water input (Figure \ref{fig:RainfallStreamflow}). During the drying phase, rainfall further back in the past begins to make a relatively larger impact on current streamflow, since the relative amount of antecedent dryness is a significant driver for streamflow production. During August (around day 230), the soil is completely dry and temperatures are high, therefore most rainfall contributes to soil wetness and evaporation instead of streamflow. Starting in mid-September (day 260), the Koksilah River catchment enters a rapid wetting phase as rainfall quickly becomes more frequent. Rainfall occurring up to around 40 days in the past begins to have a significant impact on streamflow, since during this time if the previous days were dry any additional precipitation will contribute to soil wetness, but if previous days were wet, additional precipitation will contribute to streamflow. The wetting phase ends in December (about day 350) when the soil becomes fully saturated regardless of the between-year variations of precipitation in the wettest month, November.  On the test set, $R^2 = 0.82$. 

For the second catchment, the Withlachoochee River in Florida, the $R^2$ was 0.55 on the test dataset. In the final model, the autocorrelated errors were modeled via an AR(2) process with fitted coefficients (1.389, -0.408), showing extreme autocorrelation. Though the Withlachoochee River results presented in Figure \ref{fig:WithlachoocheeResults} look very different compared to the results from the Koksilah River (Figure \ref{fig:koksilahResults}), they can be interpreted in a similar fashion. We first note that this catchment may not experience overland flow where rainfall bypasses the storage and strongly affects streamflow in a short period of time. This is reasonable since this is a dry catchment with an aridity index of greater than one, meaning that the potential for evaporation is greater than total precipitation. Starting in May (days 130-175), the catchment enters a dry phase. By this time of the year, the catchment has experienced about 6 months of relatively dry weather and temperatures have begun to rise, therefore in May, most rainfall evaporates or contributes to soil moisture. By mid to late June (about day 180), a consistent amount of rainfall usually falls, outpacing evaporation (Figure \ref{fig:RainfallStreamflow}), thus the catchment soils become wet enough to produce streamflow. Before the catchment can become fully saturated, in September (day 260), rainfall falls below evaporation and streamflow, leading to larger lags becoming significant and a drying phase from October to the next year.

Both real-world studies lead to interesting and meaningful results with good $R^2$ scores on the test set. In both cases, wetting, wet, drying, and dry periods could be delineated. Further, even though the positivity of the coefficient function was not imposed as a constraint for our models, in both cases, the model correctly identified that there is a strong positive relationship between rainfall and streamflow at all times and all lags. This is a further indication that our method is reliable and can correctly identify dominant processes. The test set $R^2$ results for both catchments are similar to those seen from neural networks which are more complex and less interpretable \citep{hoedt2021mc,anderson2022evaluation,kratzert2018rainfall}.

\section{Simulation Studies}
\label{sec:simStudies}

The aim of our simulation study is to quantify and understand our ability to estimate $\beta(s,t)$ and $\delta(t)$ in a variety of noisy environments.

\subsection{Evaluation criteria}
\label{sec:evaluationcriteria}

\citet{xun2022sparse} compared methods by computing the root mean squared error of $\delta$, the percent bias of $\delta$, and the mean integrated squared error of $\beta(s,t)$. The percent bias of $\delta$ does not fully capture the accuracy of $\delta$ estimates when $\delta$ varies over time, and the other metrics, such as the root integrated squared error, are less interpretable compared to $R^2$, thus we introduce three evaluation criteria for our simulation studies. The first criterion, $\beta$-$R^2$, aims to evaluate the accuracy in estimating the true coefficient function in an interpretable fashion:
\begin{equation}
    \label{eq:betaError}
    R^2(\beta,\hat{\beta})= 1 - \frac{\int\int\{\beta(s,t) - \hat{\beta}(s,t)\}^2\ ds \ dt}{\int\int\{\beta(s,t)- \bar{\beta}\}^2\ ds \ dt}
\end{equation}
The second criterion, $\delta(t)\text{-bias}$, evaluates the bias appearing in our estimates of $\delta(t)$. We compute this criterion as the average of $\hat{\delta}(t)$ across time minus the average of $\delta(t)$ across time. Our third criterion, $\delta(t)\text{-correlation}$ evaluates how well we can estimate the inter-period variability in $\delta(t)$ across time. This metric is simply calculated as the correlation between the ground-truth $\delta(t)$ and the estimated $\hat{\delta}(t)$. While biases could cancel over time, and strong correlation does not imply that $\hat{\delta}(t)$ has the correct scale, together these evaluation metrics account for each other's shortcomings while independently providing information about the accuracy in which we are estimating the scale and temporal dynamics of $\delta(t)$.  Ideally, $\beta$-$R^2$ and $\delta(t)\text{-correlation}$ are close to one, while $\delta(t)\text{-bias}$ is close to zero.

% For our experiments, our null hypothesis will be that $\hat{\beta}(s,t)=\Bar{\beta}$, thus $H_0(\beta)=\hat{\beta}$ and $H_0(\delta(t)=t+1$. This formulation ensures that our evaluation criteria are interpretable since a value of 1 for either of the criteria indicate and error of 0, a value of 0 indicates that the estimated parameters are no better than our null hypothesis, and a negative value indicates that the estimated parameters are worse than our null hypothesis.

\subsection{Simulation scenarios}

In this section, we run eight simulation studies that aim to quantify our method's ability to estimate $\beta(s,t)$ and $\delta(t)$ in noisy environments. We use the same real rainfall data used in the previous section to form the covariate matrix $\textbf{Z}$, then we define eight scenarios with different values for $\beta(s,t)$, $\delta(t)$, noise levels, and noise autocorrelation coefficients, such that known ground-truth values for the response $y_i(t)$ can be simulated. Streamflow is notoriously difficult to predict. Depending on the location of the catchment and its dynamics, even when using powerful black-box methods such as long short-term memory (LSTM) neural networks \citep{hochreiter1997long}, combined with more predictor variables, the test set $R^2$ can range all the way from zero to one \citep{ayzel2021effect,hoedt2021mc,kratzert2018rainfall}. In future scenarios for which our method may be applied, we hypothesize that inferences from models with $R^2<0.4$ will be difficult. Further, from previous works we know that interpretable models with $R^2>0.8$ will be rare in hydrology. Thus, we simulate the response vector with additional noise such that if we calculate the true response $y_{\mathtt{true}}(t)$ from the ground-truth $\beta(s,t)$ we would produce $R^2(y,y_{\mathtt{true}})=0.8$ or $R^2(y,y_{\mathtt{true}})=0.4$, where 
\begin{equation}
    \label{eq:maxR2}
    R^2(y,y_{\mathtt{true}})= 1 - \frac{\int{\{y(t) - y_{\mathtt{true}}(t)\}^2 \ dt}}{\int{\{y(t) - \bar{y}\}^2 \ dt}},
\end{equation}
$y(t)$ is the simulated data after adding noise to $y_{\mathtt{true}}(t)$, and $\bar{y}$ is the average of $y(t)$ over $t$. 
Further, we specify two possible levels of autocorrelation for our simulated noise. From our results in the previous section, we know that the noise can be highly autocorrelated, thus we specify medium AR(0.6,0.1) and high AR(1.5,-0.52) autocorrelation scenarios.

The eight scenarios revolve around the results obtained in the real hydrology study in Section \ref{sec:hydroStudy}. We take the results from Figures \ref{fig:koksilahResults} and \ref{fig:WithlachoocheeResults} and set them as our ground-truth scenarios. Taking these $\beta(s,t)$ values along with our matrix of rainfall values allows us to compute ground-truth response values. With two different noise levels, two different catchments, and two autocorrelation levels, there are eight simulation scenarios. For each scenario, we repeat the simulation 100 times such that stable results and uncertainty levels can be obtained. Each simulation iteration takes less than 10 minutes on an Intel i9-9980HK 2.4 GHz processor with 32 GB of RAM, so our methods are computationally feasible for any modern hardware setup.

\subsection{Simulation results}

Table \ref{tab:table} shows that the introduced method is quite promising and robust. We consistently observe that regardless of location, the data with less noise allowed for more accurate and stable inferences of $\beta(s,t)$ and $\delta(t)$. We also consistently observed that higher autocorrelation leads to poorer $\delta(t)$-corr results. The average results displayed in Table \ref{tab:table} as well as the individual simulation runs consistently resulted in negative $\delta(t)$-biases, revealing systematic errors. This indicates that on average $\delta(t) > \hat{\delta}(t)$, meaning that the estimated coefficient function is often too sparse and the chosen $q$ should decrease, suggesting that our method for choosing the threshold $q$ is not optimal. 

\begin{table}[h!]
 \caption{Summary of the simulation study results. The average (standard deviation) across the 100 replications is shown for each of the evaluation criteria defined in Section \ref{sec:evaluationcriteria}.}
  \centering
  \begin{tabular}{lcccccc}             \\
  \hline
    Location & $R^2(y,y_{\mathtt{true}})$ & AR(2) coefs & $\beta$-$R^2$ & $\delta(t)$-bias & $\delta(t)$-corr \\
    \hline
    Koksilah River & 0.4 & 0.6, 0.1 & 0.926 (0.016)  & -2.699 (1.806) & 0.772 (0.126)  \\
    Koksilah River & 0.8 & 0.6, 0.1 & 0.971 (0.005)  & -1.773 (0.769) & 0.925 (0.043)   \\
    Koksilah River & 0.4 & 1.5, -0.52 & 0.904 (0.030)  & -4.133 (2.60) & 0.534 (0.280)  \\
    Koksilah River & 0.8 & 1.5, -0.52 & 0.961 (0.009)  & -2.42 (0.944) & 0.850 (0.145)   \\
    Withlacoochee River & 0.4 & 0.6, 0.1 & 0.927 (0.014)  & -2.773 (1.684) & 0.943 (0.039)     \\
    Withlacoochee River & 0.8 & 0.6, 0.1 & 0.968 (0.005)  & -2.391 (0.571) & 0.992 (0.003)    \\
    Withlacoochee River & 0.4 & 1.5, -0.52 & 0.888 (0.037)  & -4.912 (4.155) & 0.848 (0.097)     \\
    Withlacoochee River & 0.8 & 1.5, -0.52 & 0.934 (0.015)  & -3.736 (1.045) & 0.982 (0.009)    \\
    \hline
  \end{tabular}
  \label{tab:table}
\end{table}

For the Koksilah River, the $\beta-R^2$ was consistently above 0.9 for the high noise scenarios and above 0.96 for the low noise scenarios. The high noise scenarios also gave consistently worse estimates of $\delta(t)$, with higher autocorrelation also having an impact.

For the Withlachoochee River, the estimates for $\beta(s,t)$ are consistently accurate with $R^2$s ranging from 0.888 to 0.927 in the high noise scenario and $R^2$s ranging from 0.934 to 0.968 in the low noise scenario. For both noise scenarios, the bias was about -4 days for the high correlation scenarios days while it was about -2.5 days for the low correlation scenarios. Considering $\delta(t)$ can range from 2 to over 60 (Figure \ref{fig:WithlachoocheeResults}), these bias results are quite promising. Further, the our method could recover $\delta(t)$ with a correlation above 0.98 in both low noise scenarios. The $\delta(t)$ correlation was slightly lower for the high noise low autocorrelation scenario (0.943) and much lower for the high noise high autocorrelation scenario (0.848).

\section{Discussion and conclusions}
\label{sec:conclustion}

In this work, we build off the iterations of the functional historical linear model introduced by \citet{malfait2003historical}, \citet{harezlak2007penalized}, \citet{rushworth2013distributed}, and \citet{xun2022sparse}. We greatly simplify their formulations while allowing for dynamic sparsity and high resolution estimates of the coefficient function using the Whittaker basis. In Section \ref{sec:Methods}, we illustrate our algorithm with several diagrams and introduce the three minor assumptions we make to reduce the highly overparameterized problem to one that is tractable. We only assume horizontal smoothness, vertical smoothness, and that sparsity is concentrated at larger lags to keep our methods flexible and applicable to a wide variety of applications outside of our main hydrologically-centered goal. Throughout the work, we are strongly driven by the goal of accurately estimating rainfall-runoff relationships. After gathering rainfall and streamflow data from Vancouver Island and Florida, we estimated the coefficient functions that filter rainfall into streamflow at both locations and compared our conclusions with expert knowledge. We found that the functions have a perfect hydrologically significant interpretation where catchments go through four distinct phases: (1) wetting, (2) wet, (3) drying, (4) dry. Because the Koksilah River (Vancouver Island) has high seasonal variability in temperature and rainfall, the four phases can be distinctly parsed. On the other hand, we found that the Withlachoochee River (Florida) is an extremely dry catchment, therefore it never enters the wet phase, but all other phases can be inferred. Indeed, none of these inferences could have been made without the methodological improvements pertaining to the dynamic sparsity. Finally, in an extensive simulation study, we found that our simple methodology can successfully recover the ground-truth coefficient function $\beta(s,t)$ as well as the dynamic time lag $\delta(t)$ with high accuracy. Though the accuracy for which we can recover the coefficient function and time-lag clearly depends on the level of noise and autocorrelation, we found that even when the $R^2$ is less than or equal to 0.4 and the errors are extremely autocorrelated, accurate conclusions can be made from our inferences.

Several shortcomings of our methods and experiments can be identified. First, though our algorithm can potentially include multiple features, these additional features may not follow our sparsity and smoothness assumptions. For example, if we included snowfall as a predictor of streamflow in a snow-rich catchment, we may expect snowfall to take days or weeks or even months to melt and begin to contribute to streamflow. This would contradict our sparsity assumption since very low and very high lags would be expected to have zero coefficients. Second, we assume throughout the work that the behavior of the catchment does not change across different years. Although our method does not capture year-to-year relationship variability, our method can be used to extract multiple coefficient functions before and after some known change. For example, one could estimate the rainfall-runoff relationship before and after a large change in land use or before and after a period of severe climate change. Focusing now on the limitations of our experiments, we note that our experiments are only limited to a specific application in hydrology, though we hypothesize that our methods can work well in other domains. Further, our simulation study is limited since we provided ground-truths that we know can be reached from the algorithm. This limitation points back to the second methodological limitation since our method may only be able to recover certain types of coefficient functions. If we instead ran a simulation study with a non-smooth ground-truth, the simulation results would likely be worse.

In future work, one could solve one or more of the above shortcomings. Perhaps the flexibility of our methods could also be improved by adding adaptive smoothness in a similar fashion as the adaptive lasso, since different days or lags could have different smoothness \citep{zou2006adaptive,centofanti2022adaptive,yang2017adaptive,martinez2010oracle,ballout2023use}. Adding time-varying AR terms to the autoregressive distributed lag model or accounting for heterscedasticity may also improve the extraction of filtering functions \citep{kim2023time,rao2004multiple}. Alternatively, extending these methods to spatial analysis could be interesting. With smooth spatial attribution, one could understand how impactful a specific grid cell of a cause is on the response. Further, with sparsity, one could delineate the spatial area that is impactful on the response versus the areas which have no significant impact on the response. Our current work takes the perspective of penalized least squares and sparsity thresholding, but another interesting perspective leading to different methods could come from a causal inference and conditional independence framework \citep{laumann2023kernel}. For example, finding the correct sparsity using partial correlation or transfer entropy may be possible \citep{murari2018use,moges2022strength}, though the high dimensional and low sample size environment makes this non-trivial. Better methods for choosing our hyperparameters could also be explored \citep{bach2008bolasso,hall2009bootstrap,chatterjee2011bootstrapping}. We hope our methods can serve as a reliable tool for learning from time series data across multiple areas of science or serve as an inspiration for further methodological development. This area of research is certainly rich with potential scientific discoveries.

\bigskip
\begin{center}
{\large\bf SUPPLEMENTARY MATERIAL}
\end{center}

\begin{description}

\item[R code and for running experiments:] The R code file jasa\_code\_Submitted.R contains all the code needed to run the experiments. Further, the four .rds files contain all streamflow and rainfall data. See \url{https://github.com/HydroML/HFLM-DS} for all files.

\end{description}

% \section{BibTeX}

% We hope you've chosen to use BibTeX!\ If you have, please feel free to use the package natbib with any bibliography style you're comfortable with. The .bst file agsm has been included here for your convenience. 

\bibliographystyle{apalike}

\bibliography{references.bib}
\end{document}